\documentclass[onecolumn,amsmath,amssymb,superscriptaddress,nofootinbib,11pt,a4paper]{revtex4}

\pdfoutput=1
\usepackage[T1]{fontenc}
\usepackage{CJK}
\usepackage{xcolor}
\usepackage{amsfonts}
\usepackage{epstopdf}
\usepackage{wrapfig} % to wrap figure with
\usepackage{subcaption}
\usepackage{graphicx}  % Include figure files
\usepackage{dcolumn}   % Align table columns
\usepackage{bm}
\usepackage{float}
\usepackage[T1]{fontenc}
\usepackage{CJK}
\usepackage{xcolor}
\usepackage{amsfonts}
\usepackage{epstopdf}
\usepackage{wrapfig} % to wrap figure with
\usepackage{subcaption} 
\usepackage{graphicx}  % Include figure files
\usepackage{dcolumn}   % Align table columns
\usepackage{bm}
\usepackage{float}
\usepackage[utf8]{inputenc}
\usepackage[T1]{fontenc}

\begin{document}

\title{Repetitive Penrose Process in Accelerating Kerr Black Holes}

\author{Xiao-Xiong Zeng}
\email{xxzengphysics@163.com}
\affiliation{College of Physics and Electronic Engineering, Chongqing Normal University, \\Chongqing 401331, China}

\author{Ke Wang\footnote{Electronic address: kkwwang2025@163.com  (Corresponding author)}}
\affiliation{School of Material Science and Engineering, Chongqing Jiaotong University, \\Chongqing 400074, China}
\begin{abstract}
{ This paper investigates the repetitive Penrose process in accelerating Kerr black holes and explores the influence of the acceleration factor on the repetitive Penrose process. After a brief review of accelerating Kerr black holes, we study the fundamental equations of the Penrose process in this spacetime, examine the stopping conditions required for the repetitive Penrose process, and obtain corresponding numerical results. The conclusions indicate that, in addition to exhibiting previously observed similar phenomena, accelerating Kerr black holes exhibit stronger energy extraction capabilities compared to Kerr black holes during the repetitive Penrose process. Moreover, in prior studies, the energy utilization efficiency was difficult to exceed $50\%$. However, in accelerating Kerr black holes, when the decay radius is relatively small, the energy utilization efficiency can exceed $50\%$, indicating that the reduced extractable energy primarily transforms into extracted energy rather than irreducible mass. On the other hand, when the initial value of the acceleration factor is large, the extractable energy can decrease to nearly zero, which also differs from the case of Kerr black holes in previous studies.
}
\end{abstract}

\maketitle
%\flushbottom
% \tableofcontents
 \newpage
\section{Introduction}
The Penrose process is a mechanism for extracting energy from a rotating black hole \cite{8}. In this process, a particle falls from infinity and decays within the ergosphere. One of the resulting particles carries negative energy and ultimately falls into the event horizon. Due to energy conservation, the other particle escapes to infinity carrying more energy than the original particle, which is equivalent to extracting rotational energy from the black hole. Since its proposal, this process has garnered widespread attention from physicists \cite{28,29,30,31,32,33,34,35,36,37}. Research indicates that for energy extraction to be feasible, the velocity of the outgoing particle should exceed half the speed of light \cite{9,10}.

Penrose's pioneering work has inspired physicists to explore various alternative mechanisms for extracting energy from black holes \cite{25,11,12,13,14,15,16,17,26,18}. Recently, Ruffini et al. \cite{19} achieved energy extraction by imposing a turning point condition on the particle's trajectory within the original equations of motion for the Penrose process in Kerr black holes. Furthermore, they proposed a repetitive Penrose process \cite{3} (for related early work, see \cite{27}). Research found that in the repetitive Penrose process, it is impossible to extract all the extractable energy from a Kerr black hole, the change in extractable energy primarily transforms into irreducible mass. This is because after each Penrose process, the new black hole mass and spin must be used for subsequent energy extraction, while also accounting for the new irreducible mass. The repetitive Penrose process is nonlinear, and the irreducible mass increases nonlinearly as well. The repetitive Penrose process proposed by Ruffini has since been extended to the repetitive electric Penrose process \cite{6} and to Kerr-de Sitter black holes \cite{4}. In the repetitive electric Penrose process, similarly, the total electric energy of a Reissner-Nordström black hole cannot be fully depleted. In Kerr-de Sitter black holes, besides exhibiting phenomena similar to previous cases, the energy extraction capability in certain scenarios is stronger than that of Kerr black holes.

The C-metric is a special class of black hole solutions in general relativity, which can describe accelerating black holes \cite{20} and belongs to the well-known Plebański–Demiański spacetime \cite{21}. Accelerating black holes typically refer to black holes moving with acceleration along a certain direction under specific conditions, such as being influenced by extreme gravitational sources like cosmic strings. The inclusion of an acceleration factor fundamentally alters the causal structure and asymptotic properties of the black hole spacetime. Currently, many properties of accelerating black holes have been extensively studied \cite{22,23,24,1,2}. In this paper, we aim to further investigate the repetitive Penrose process in accelerating Kerr black holes and explore the influence of the acceleration factor on this process. In particular, we conduct a detailed study of several iterative stopping conditions. The results indicate that, in addition to the analogous characteristics previously noted, accelerating Kerr black holes possess stronger energy extraction capabilities compared to Kerr black holes during the repetitive Penrose process. Moreover, in prior research, the energy utilization efficiency was difficult to exceed $50\%$. However, in accelerating Kerr black holes, when the decay radius is relatively low, the energy utilization efficiency can exceed $50\%$, indicating that the reduced extractable energy primarily transforms into extracted energy rather than irreducible mass. On the other hand, when the initial value of the acceleration factor is large, the extractable energy can decrease to nearly zero, which also differs from previous conclusions. These findings demonstrate that conducting the repetitive Penrose process in accelerating Kerr black holes is easier and yields more extracted energy.

The remainder of this paper is organized as follows. In Section 2, we will introduce the Penrose process in accelerating Kerr black holes. In Section 3, we present the iterative stopping conditions for this process. In Section 4, we will study the repetitive Penrose process in accelerating Kerr black holes, with a particular focus on exploring the influence of the acceleration factor on this process. We conclude in Section 5. Throughout the paper, we will adopt natural units $(c=G=1)$.

\section{The Penrose Process in Accelerating Kerr Black Holes}
In the Boyer-Lindquist coordinates, the metric for an accelerating Kerr black hole is given by \cite{1,2}
\begin{equation}
\begin{split}
ds^2 = \frac{1}{H^2} \Biggl[ & \frac{1}{\alpha^2 \Sigma} \left( -\Delta_r + a^2 \Delta_\theta \sin^2\theta \right) dt^2 + \frac{2a\sin^2\theta}{\alpha \Sigma} \left( \Delta_r  - \Delta_\theta (r^2 + a^2) \right) dt d\phi \\
& + \frac{\Sigma}{\Delta_r} dr^2 + \frac{\Sigma}{\Delta_\theta} d\theta^2 + \frac{\sin^2\theta}{\Sigma} \left( -a^2 \Delta_r \sin^2\theta + \Delta_\theta (r^2 + a^2)^2 \right) d\phi^2 \Biggr],
\end{split}
\end{equation}
where
\begin{equation}
\begin{aligned}
H &= 1 + Ar \cos \theta,  \Sigma = r^2 + a^2 \cos^2 \theta, \Delta_r = (1 - A^2 r^2)(r^2 - 2Mr + a^2 ),\\
\Delta_\theta &= 1 + 2MA \cos \theta + A^2a^2\cos^2 \theta,\alpha = \sqrt{\frac{1 - a^2 A^2}{1 + a^2 A^2}}.
\end{aligned}
\end{equation}
Here, $M$ is the black hole mass, $a$ is the black hole spin, and $A$ is the acceleration factor. If $A=0$, the metric reduces to the Kerr metric. 

The event horizon of the black hole is determined by $\Delta_r=0$, i.e.,
\begin{equation}
r_A = \dfrac{1}{A},r_\pm = M \pm\sqrt{M^{2} - a^{2}}.
\end{equation}
Among these, $r_A$ is the acceleration horizon, while $r_+$ and $r_-$ represent the event horizon and the Cauchy horizon, respectively. In general, we have $r_- \leq r_+ < r_A$. In this paper, to maximize energy extraction, we adopt a simplifying assumption by considering the repetitive Penrose process for an extremal accelerating Kerr black hole on the equatorial plane, where the initial black hole spin is $a=M$. Likewise, only the initial black hole is extremal; it does not remain extremal after the iterations. Indeed, it is impossible for the black hole to remain extremal throughout the repetitive Penrose process. This simplification has been widely employed in references \cite{3,4}.

The boundary of the black hole's ergosphere is determined by $g_{tt}=0$. On the equatorial plane, the solution to this equation is
\begin{equation}
\begin{cases}
r_{E0} = 0, \\[2ex]
r_{E-} = \dfrac{1}{3} \left( -\dfrac{\sqrt[3]{Q}}{A^2} + \dfrac{A^2 (3 a^2 - 4 M^2) - 3}{\sqrt[3]{Q}} + 2 M \right), \\[2ex]
r_{EA} = \dfrac{1}{6} \left( \dfrac{(1 - i \sqrt{3}) \sqrt[3]{Q}}{A^2} - \dfrac{(1 + i \sqrt{3}) \left( A^2 (3 a^2 - 4 M^2) - 3 \right)}{\sqrt[3]{Q}} + 4 M \right), \\[2ex]
r_{E} = \dfrac{1}{6} \left( \dfrac{(1 + i \sqrt{3}) \sqrt[3]{Q}}{A^2} - \dfrac{(1 - i \sqrt{3}) \left( A^2 (3 a^2 - 4 M^2) - 3 \right)}{\sqrt[3]{Q}} + 4 M \right),
\end{cases}   
\end{equation}
where
\begin{equation}
\begin{cases}
Q = A^6 (9 a^2 M - 8 M^3) + 3 \sqrt{3} \sqrt{D} + 18 A^4 M,\\[2ex]
D = A^6 \left( (a^2 A^2 - 1)^3 + A^2 M^2 (-a^4 A^4 + 20 a^2 A^2 + 8) - 16 A^4 M^4 \right).
\end{cases}   
\end{equation}
Here, $r_{E-} < 0$ and is therefore discarded. $r_{E}$ is the ergosphere boundary of the event horizon, whose value is greater than the event horizon, and $r_{EA}$ is the ergosphere boundary of the acceleration horizon, whose value is less than the acceleration horizon. In this paper, we extract energy from  the ergosphere of the event horizon. Typically, $r_{E} < r_{EA}$. When the acceleration factor increases to a certain value, $r_{E}$ and $r_{EA}$ become equal. Beyond this value, the ergosphere ceases to exist, making energy extraction impossible. For example, for an extremal black hole with $a=M$, the acceleration factor $\hat{A}=AM = \sqrt{\frac{5 \sqrt{5}}{2} - \frac{11}{2}} = 0.30028$, at which point $r_{E} = r_{EA} = \frac{1}{2} (\sqrt{5}+3) M = 2.61803M$. Here, a hat over a symbol denotes a dimensionless quantity.

\begin{figure}[!h]
  \centering
  \setlength{\tabcolsep}{2pt}
  \begin{tabular}{ccc}
    \includegraphics[width=0.45\linewidth]{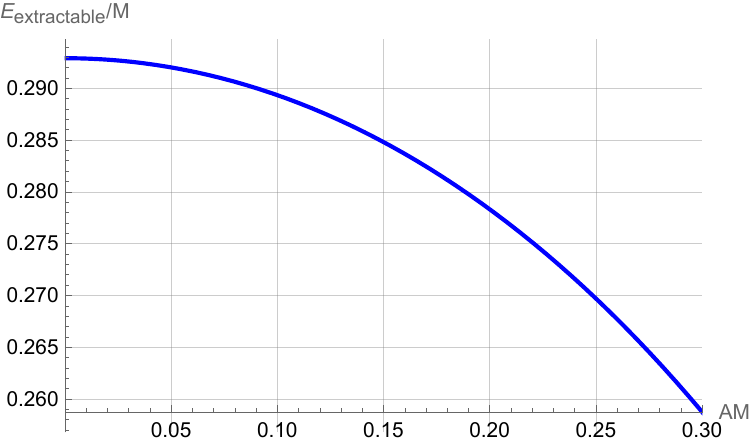} &
    \includegraphics[width=0.45\linewidth]{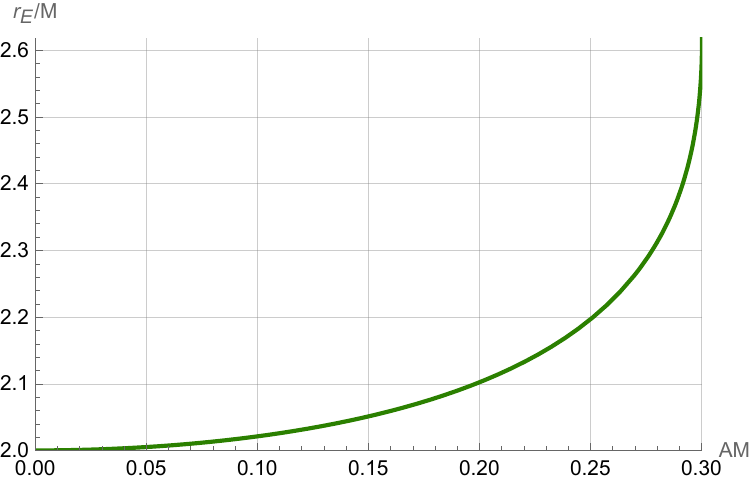} 
  \end{tabular}
  \caption{Variation of $E_{extractable}/M$ and $r_{E}/M$ with $\hat{A}$ for an extremal black hole.}
  \label{fig:1}
\end{figure}

The surface area of the black hole's event horizon is
\begin{equation}
S = \int_0^\pi \int_0^{2\pi} \sqrt{g_{\theta\theta} g_{\phi\phi}} \, d\phi d\theta = \frac{4\pi (r_+^2 + a^2)}{1 - A^2 r_+^2}.
\end{equation}
The irreducible mass of the black hole is \cite{5}
\begin{equation}
M_{irr} = \sqrt{\frac{S}{16\pi}}=\sqrt{\frac{r_+^2 + a^2}{4 (1 - A^2 r_+^2)}}.
\end{equation}
Thus, the extractable energy can be expressed as
\begin{equation}
E_{extractable}= M-M_{irr}=M-\sqrt{\frac{r_+^2 + a^2}{4 (1 - A^2 r_+^2)}}.
\end{equation}
This represents the theoretically maximum extractable energy that can be extracted from an accelerating Kerr black hole. For an extremal black hole, the irreducible mass and the extractable energy are
\begin{equation}
M_{irr,0} = \frac{M}{\sqrt{2(1 - A^2 M^2)}},E_{extractable,0}= M-\frac{M}{\sqrt{2(1 - A^2 M^2)}}.
\end{equation}
Therefore, as $\hat{A}$ increases, the maximum extractable energy decreases. In Fig. \ref{fig:1}, we plot the maximum extractable energy and the location of the ergosphere boundary at the event horizon for an extremal black hole as functions of $AM$.
It can be observed that as $\hat{A}$ increases, the ergosphere boundary at the event horizon of the extremal black hole gradually increases, while the maximum extractable energy gradually decreases.

The fundamental equations of the Penrose process are given by 4-momentum conservation. The conservation equations for energy, angular momentum, and radial momentum can be expressed as
\begin{equation}
\begin{cases}
\hat{E}_0=\tilde{\mu}_1\hat{E}_1+\tilde{\mu}_2\hat{E}_2, \\[2ex]
\hat{p}_{\phi 0}=\tilde{\mu}_1\hat{p}_{\phi 1}+\tilde{\mu}_2\hat{p}_{\phi 2},\\[2ex]
\hat{p}_{r0}=\tilde{\mu}_1\hat{p}_{r1}+\tilde{\mu}_2\hat{p}_{r2},
\end{cases}\label{10}
\end{equation}
where
\begin{equation}
\hat{E}_i={E}_i/{\mu}_i,\hat{p}_{\phi i}=p_{\phi i}/(\mu_iM),\hat{p}_{ri}=p_{ri}/\mu_i,\tilde{\mu}_i=\mu_i/\mu_0,i\in0,1,2.
\end{equation}
Here, $\mu_{i}$ is the mass of particle $i$. The effective potential for the radial motion of particles on the equatorial plane is given by \cite{4}
\begin{equation}
\hat{V}^\pm_i = \frac{ g^{t\phi} \hat{p}_{\phi i} M \mp \sqrt{ (g^{t\phi})^2 \hat{p}_{\phi i}^2 M^2 - g^{tt} ( g^{\phi\phi} \hat{p}_{\phi i}^2 M^2 + 1 ) } }{ g^{tt} }.
\end{equation}
We focus on the optimal conditions for maximum energy extraction, where the radial momentum of all three particles must be zero at the decay position. This means that the three particles locates at their respective turning points, i.e., $\hat{E}_i = \hat{V}_i^+$. Specific reasons for this can be found in reference \cite{4}. Under this condition, assuming $\hat{E}_0$, $\hat{p}_{\phi 1}$, and $\nu = \mu_2/\mu_1$ are known quantities, the fundamental equations of the Penrose process \eqref{10} have an analytical solution \cite{4}
\begin{equation}
\begin{cases}
\hat{p}_{\phi 0} &=\frac{ g^{t\phi} \hat{E}_0 +\sqrt{ (g^{t\phi})^2 \hat{E}_0^2 - g^{\phi\phi} (1 + g^{tt} \hat{E}_0^2) } }{ M g^{\phi\phi} }, \\[2ex]
\hat{E}_1 &= \frac{ g^{t\phi} \hat{p}_{\phi 1} M - \sqrt{ (g^{t\phi})^2 \hat{p}_{\phi 1}^2 M^2 - g^{tt} ( g^{\phi\phi} \hat{p}_{\phi 1}^2 M^2 + 1 ) } }{ g^{tt} },\\[2ex]
\tilde{\mu}_1&= \frac{\hat{E}_0\hat{E}_1g^{tt} - \hat{E}_1g^{t\phi}M\hat{p}_{\phi 0} - \hat{E}_0g^{t\phi}M\hat{p}_{\phi 1} + g^{\phi\phi}M^2\hat{p}_{\phi 0}\hat{p}_{\phi 1} + \sqrt{F}}{\hat{E}_1^2g^{tt} - 2\hat{E}_1g^{t\phi}M\hat{p}_{\phi 1} + g^{\phi\phi}M^2\hat{p}_{\phi 1}^2 + \nu^2},\\[2ex]
\hat{E}_2 &=\frac{\hat{E}_0}{\tilde{\mu}_2}-\frac{\hat{E}_1}{\nu}, \hat{p}_{\phi 2}=\frac{\hat{p}_{\phi 0}}{\tilde{\mu}_2}-\frac{\hat{p}_{\phi 1}}{\nu},  \label{13} 
\end{cases}
\end{equation}
where
\begin{equation}
\begin{aligned}
F= & - g^{tt} g^{\phi\phi} M^2 \hat{E}_1^2 \hat{p}_{\phi 0}^2  + (g^{t\phi})^2 M^2 \hat{E}_1^2 \hat{p}_{\phi 0}^2  - g^{tt} g^{\phi\phi} M^2 \hat{E}_0^2 \hat{p}_{\phi 1}^2  + (g^{t\phi})^2 M^2 \hat{E}_0^2 \hat{p}_{\phi 1}^2 \\
& + 2 g^{tt} g^{\phi\phi} M^2 \hat{E}_0 \hat{E}_1 \hat{p}_{\phi 0} \hat{p}_{\phi 1}  - 2 (g^{t\phi})^2 M^2 \hat{E}_0 \hat{E}_1 \hat{p}_{\phi 0} \hat{p}_{\phi 1}  - g^{tt} \hat{E}_0^2 \nu^2  + 2 g^{t\phi} M \hat{E}_0 \hat{p}_{\phi 0} \nu^2 \\
& - g^{\phi\phi} M^2 \hat{p}_{\phi 0}^2 \nu^2.
\end{aligned}
\end{equation}
After each energy extraction, the remaining mass and angular momentum of the black hole are given by
\begin{equation}
M_n=M_{n-1}+\hat{E}_{1,n-1}\mu_{1,n-1},L_n=L_{n-1}+\hat{p}_{\phi 1}\mu_{1,n-1}M_{n-1},
\end{equation}
where
\begin{equation}
L_0=\hat{a}_0M_0^2=M_0^2.
\end{equation}
This leads to corresponding changes in $\hat{a} = a/M$ and $\hat{A}$, namely
\begin{equation}
\Delta\hat{a}_{n-1}=\frac{L_n}{M_n^2}-\frac{L_{n-1}}{M_{n-1}^2},\Delta\hat{A}_{n-1}=A M_n-A M_{n-1}.
\end{equation}
The change in the event horizon is
\begin{equation}
\Delta r_{+,n-1}=M_{n}\left(1+\sqrt{1-\hat{a}_{n}^{2}}\right)-M_{0}\left(1+\sqrt{1-\hat{a}_{0}^{2}}\right).
\end{equation}
The change in irreducible mass is
\begin{equation}
\Delta M_{ irr,n-1}=\sqrt{\frac{r_{+,n}^2 + (\hat{a}_{n}M_n)^2}{4 (1 - A^2 r_{+,n}^2)}}-\sqrt{\frac{r_{+,0}^2 + (\hat{a}_{0}M_0)^2}{4 (1 - A^2 r_{+,0}^2)}}.
\end{equation}
The change in extractable energy is
\begin{equation}
\Delta E_{ extractable,{n-1}}=\Delta M_{n-1}-\Delta M_{ irr,{n-1}}.
\end{equation}
During this process, the extracted energy is
\begin{equation}
E_{ extracted,n}=M_{0}-M_{n}.
\end{equation}
The energy return on investment, defined as the ratio of extracted energy to the total energy of all incident particles from infinity, is given by \cite{6}
\begin{equation}
\xi_{n}=E_{ extracted,n}/(nE_{0}).
\end{equation}
The energy utilization efficiency, defined as the ratio of extracted energy to the difference between the initial and final extractable energies, is given by \cite{6}
\begin{equation}
\Xi_n=E_{ extracted,n}/(E_{ extractable,0}-E_{ extractable,n}).
\end{equation}

The above formulas will serve as important parameters for evaluating the strength of energy extraction and will be discussed in more detail later.

\section{Iterative Stopping Conditions}
\begin{figure}[!h]
  \centering
  \setlength{\tabcolsep}{2pt}
  \begin{tabular}{ccc}
    \subcaptionbox{Particle 0\label{fig:2a}}[0.32\linewidth]{\includegraphics[width=\linewidth]{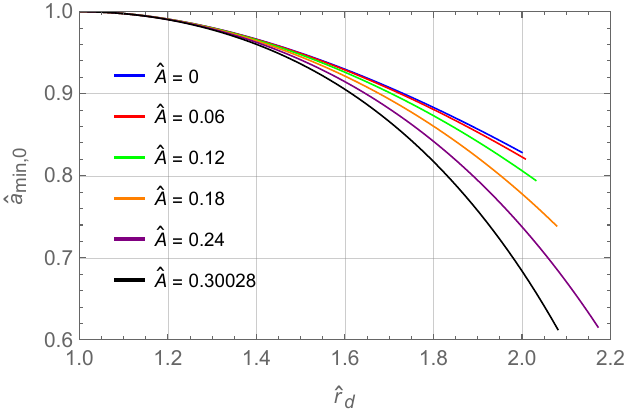}} &
    \subcaptionbox{Particle 1\label{fig:2b}}[0.32\linewidth]{\includegraphics[width=\linewidth]{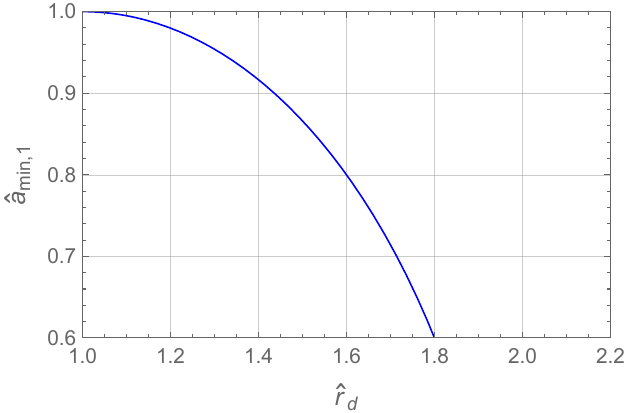}} &
    \subcaptionbox{Particle 2\label{fig:2c}}[0.32\linewidth]{\includegraphics[width=\linewidth]{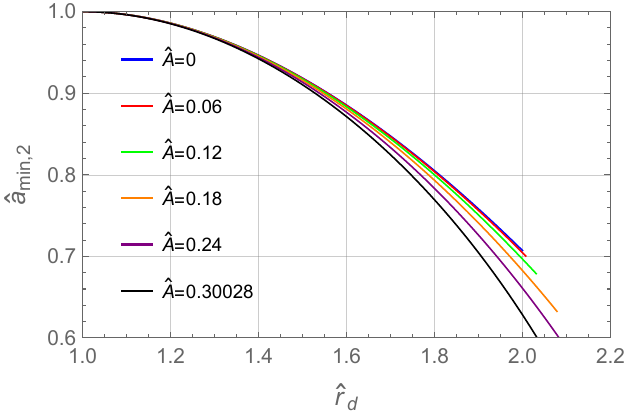}}
  \end{tabular}
  \caption{Variation of the lower spin limit with decay radius $\hat r_d$ for different $\hat{A}$ values for (a) particle 0, (b) particle 1, and (c) particle 2.}
  \label{fig:2}
\end{figure}

During the repetitive energy extraction process, the aforementioned iteration cannot proceed indefinitely and must satisfy the following five conditions. First, the mass deficit must satisfy
\begin{equation}
1-\tilde{\mu}_1-\tilde{\mu}_2>0.\label{24}
\end{equation}
This condition is easily met if the parameters are appropriately chosen. Second, during the iteration, it is required that $\hat{E}_{1}<0$. Third, for each iteration, it must hold that $E_{ extractable,n}>0$. Fourth, for each iteration, the irreducible mass must not decrease. Since the irreducible mass remains constant under reversible transformations and increases under irreversible transformations, a decrease in irreducible mass is strictly prohibited, as this would correspond to a decrease in entropy. Finally, the turning points of particle 0 and particle 2 must lie to the right of the peak of their effective potentials, whereas the turning point of particle 1 must lie to the left of the peak of its effective potential. The corresponding limiting case occurs when the classical turning point of each particle coincides exactly with the peak of its respective effective potential, i.e.,
\begin{equation}
\hat{V}_i^{+}(\hat{r}_d)=\hat{E}_i, {d}\hat{V}_i^+/{ d}\hat{r}|_{\hat{r}=\hat{r}_d}=0,
\end{equation}
where $\hat{r}_d = {r}_d/M$ is the dimensionless decay radius. If $\hat{E}_0=1$, the lower spin limit for stopping the iteration is governed by particle 0, with its lower spin limit located at the corotating marginally bound orbit of that particle. The angular velocity for corotating Keplerian orbital motion in an accelerating Kerr black hole is given by \cite{7}
\begin{equation}
\Omega_K = \frac{-\partial_r g_{t\phi} + \sqrt{(\partial_r g_{t\phi})^2 - (\partial_r g_{tt})(\partial_r g_{\phi\phi})}}{\partial_r g_{\phi\phi}}=\frac{ \sqrt{ M(1 + A^2 r^2) - A^2 r^3 } }{ \alpha \left( r^{3/2} + a \sqrt{ M(1 + A^2 r^2) - A^2 r^3 } \right) }.
\end{equation}
The corresponding specific energy for corotating Keplerian orbital motion is \cite{7}
\begin{equation}
\hat{\mathcal{E}} = -\frac{g_{tt} + g_{t\phi} \Omega_K}{\sqrt{-g_{tt} - 2g_{t\phi} \Omega_K - g_{\phi\phi} \Omega_K^2}}=\frac{ (\Delta_r - a^2) + a \sqrt{r} X }{ \alpha r \sqrt{ \Delta_r - a^2 + 2a \sqrt{r} X - r X^2 } },\label{27}
\end{equation}
where
\begin{equation}
X = \sqrt{ M(1 + A^2 r^2) - A^2 r^3 }.
\end{equation}
For a marginally bound orbit, $\hat{\mathcal{E}}=1$. The lower spin limit for particle 0 can then be obtained by solving equation \eqref{27}. If $\hat{E}_0>1$, the lower spin limit for stopping the iteration is governed by particle 2, with its lower spin limit located at the corotating photon sphere radius. The corotating photon sphere radius in an accelerating Kerr black hole satisfies \cite{1}
\begin{equation}
(4r\Delta_{r}-\Sigma\partial_{r}\Delta_{r})^{2}= 16a^{2}r^{2}\Delta_{r}\Delta_{\theta}\sin^{2}\theta.\label{29}
\end{equation}
Thus, the lower spin limit for particle 2 can be obtained by solving equation \eqref{29}. For the turning point of particle 1 to exist, the discriminant under the square root in the second formula of equation \eqref{13} must be positive, which corresponds to $\hat r_d$ being greater than the radius of the black hole's event horizon. The critical case occurs when $\hat r_d = \hat r_+$. Therefore, the lower spin limit for particle 1 is
\begin{equation}
\hat{a}_{min,1}=\sqrt{\hat{r}_d(2-\hat{r}_d)}.
\end{equation}
That is, the lower spin limit for particle 1 is independent of $\hat{A}$. In Fig. \ref{fig:2}, we plot the variation of the lower spin limits for the three particles with $\hat r_d$ under different $\hat{A}$ values. For different $\hat{A}$ values, the ergosphere varies, leading to different allowable ranges for the decay radius. Furthermore, we have omitted the regions with large $\hat r_d$ and very small $\hat{a}_{min}$, which does not affect the overall trend.
From Fig. \ref{fig:2}, it can be observed that as the decay radius increases, the lower spin limits for all three particles decrease. As $\hat{A}$ increases, the lower spin limits for particle 0 and particle 2 decrease. This is a positive sign because a lower minimum spin indicates a greater number of possible iterations, thereby allowing more energy to be extracted during the repetitive Penrose process. Furthermore, unlike the Kerr-de Sitter spacetime where the lower spin limits for all three particles increase with the cosmological parameter $\hat{\Lambda}$ \cite{4}, the accelerating Kerr spacetime exhibits the opposite trend.

\begin{figure}[!h]
  \centering
  \setlength{\tabcolsep}{2pt}
  \begin{tabular}{ccc}
    \subcaptionbox{$\hat{A}=0$}[0.32\linewidth]{\includegraphics[width=\linewidth]{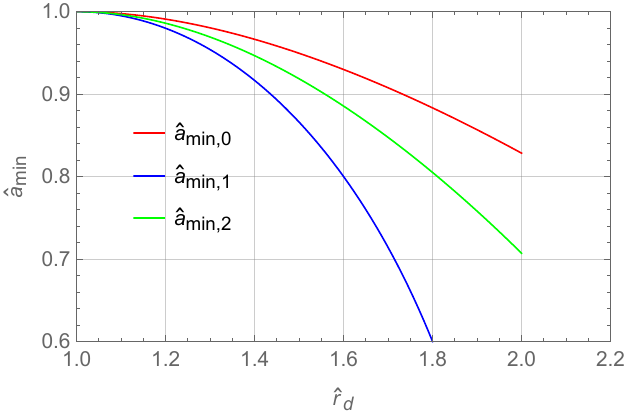}} &
    \subcaptionbox{$\hat{A}=0.18$}[0.32\linewidth]{\includegraphics[width=\linewidth]{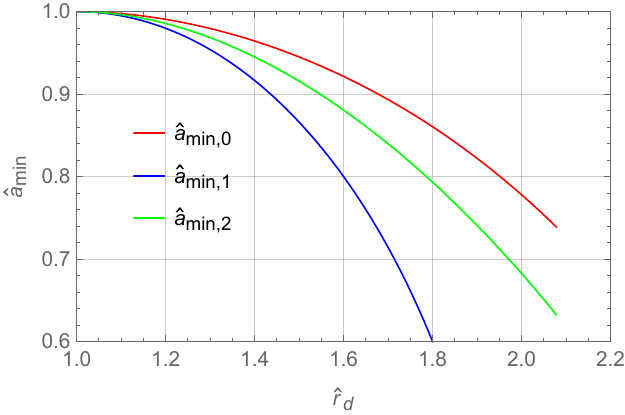}} &
    \subcaptionbox{$\hat{A}=0.30028$}[0.32\linewidth]{\includegraphics[width=\linewidth]{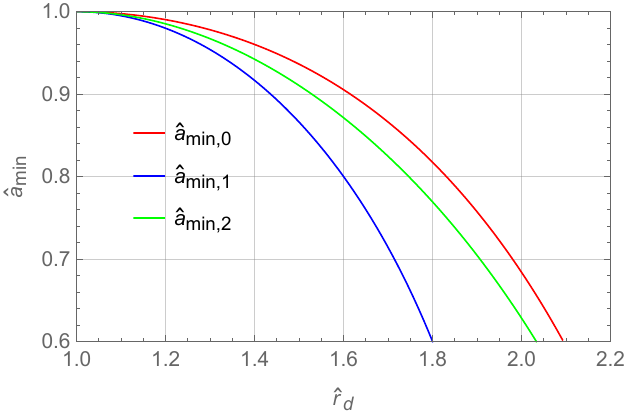}}
  \end{tabular}
  \caption{Comparison of the lower spin limits for the three particles.}
  \label{fig:3}
\end{figure}

Next, in order to determine which of particles 0, 1, and 2 actually governs the lower spin limit for stopping the iteration, we plot Fig. \ref{fig:3}.
From the figure, it can be observed that $\hat{a}_{min,1}<\hat{a}_{min,2}<\hat{a}_{min,0}$. Therefore, the lower spin limit for stopping the iteration is governed by particle 0. It is important to emphasize that this lower spin limit is not fixed. Because after each iteration, the mass $M$ decreases, leading to a corresponding decrease in $AM$. Consequently, as inferred from Fig. \ref{fig:2a}, the lower spin limit for stopping the iteration slightly increases with each iteration, but it will not exceed that of a Kerr black hole. The variation of the effective potential for particle 0 with $\hat{r}$ is plotted in Fig. 2 of Ref. \cite{3}, Fig. 6 of Ref. \cite{4}, and Fig. 2 of Ref. \cite{6}, further explaining why the lower spin limit for iteration stopping is controlled by particle 0.

\section{Repetitive Penrose Process in Accelerating Kerr Black Holes}
In this section, we study the repetitive Penrose process. The results in reference \cite{3} show that for cases with $\hat{E}_0>1$, the energy return on investment is lower than the case with $\hat{E}_0=1$. Therefore, we choose $\hat{E}_0=1$ to maximize the energy return on investment. Following reference \cite{3}, we set $\hat{p}_{\phi 1}=-19.434$, $\nu = \mu_2/\mu_1=0.78345$, and take $\mu_0 = 10^{-2}M$. We present the results in Table \ref{tab:1}. The initial dimensionless acceleration factor is taken as $\hat{A}=0.3$, in which case the ergosphere is located at $\hat{r}\in(1,2.5726)$. We choose the decay radius as $\hat{r}_d=1.2$.
\begin{table}[htbp]
\centering
\small
\caption{Repetitive Penrose process with initial $\hat{A}=0.3$ and $\hat{r}_d=1.2$.}
\label{tab:1}
\begin{tabular}{ccccccccccc}
\hline
$n$ & $\frac{M_n}{M_0}$ & $\hat{a}_n$ & $\frac{\mu_{1,n}}{\mu_0}$ & $\hat{E}_{1,n}$ & $\frac{E_{extractable,n}}{M_0}$ & $\frac{E_{extracted,n}}{M_0}$ & $\frac{M_{{irr,n}}}{M_0}$ & $\xi_n$ & $\Xi_n$ &$\hat{a}_{min,0,n}$ \\
\hline
0 & 1.000000 & 1.000000 & 0.022251 & -7.672332 & 0.258751 & 0.000000 & 0.741249 & 0.000000 & 0.000000 & 0.990248 \\
1 & 0.998293 & 0.999084 & 0.022173 & -7.691577 & 0.239495 & 0.001707 & 0.758798 & 0.170718 & 0.088659 & 0.990250 \\
2 & 0.996587 & 0.998175 & 0.022093 & -7.711233 & 0.231461 & 0.003413 & 0.765127 & 0.170630 & 0.125049 & 0.990253 \\
3 & 0.994884 & 0.997274 & 0.022014 & -7.731321 & 0.225307 & 0.005116 & 0.769577 & 0.170542 & 0.152982 & 0.990255 \\
4 & 0.993182 & 0.996380 & 0.021933 & -7.751866 & 0.220141 & 0.006818 & 0.773041 & 0.170455 & 0.176594 & 0.990257 \\
5 & 0.991482 & 0.995493 & 0.021852 & -7.772896 & 0.215616 & 0.008518 & 0.775866 & 0.170369 & 0.197484 & 0.990260 \\
6 & 0.989783 & 0.994615 & 0.021770 & -7.794440 & 0.211552 & 0.010217 & 0.778231 & 0.170283 & 0.216468 & 0.990262 \\
7 & 0.988086 & 0.993745 & 0.021687 & -7.816535 & 0.207844 & 0.011914 & 0.780243 & 0.170197 & 0.234030 & 0.990265 \\
8 & 0.986391 & 0.992883 & 0.021604 & -7.839220 & 0.204420 & 0.013609 & 0.781971 & 0.170112 & 0.250487 & 0.990267 \\
9 & 0.984697 & 0.992030 & 0.021519 & -7.862540 & 0.201234 & 0.015303 & 0.783463 & 0.170028 & 0.266055 & 0.990269 \\
10 & 0.983006 & 0.991187 & 0.021433 & -7.886547 & 0.198250 & 0.016995 & 0.784756 & 0.169945 & 0.280897 & 0.990272 \\
11 & 0.981315 & 0.990352 & 0.021346 & -7.911304 & 0.195441 & 0.018685 & 0.785875 & 0.169862 & 0.295131 & 0.990274 \\
\hline
\end{tabular}
\end{table}

All data in Table \ref{tab:1} satisfy the iterative conditions. For example, according to the mass deficit formula \eqref{24}, we obtain $\tilde{\mu}_{1} < 1/(1+\mu_{2}/\mu_{1}) = 0.56$. Additionally, for each iteration, the conditions $\hat{E}_{1} < 0$ and $E_{extractable,n} > 0$ are also satisfied, and the irreducible mass does not decrease. The last column represents the lower spin limit for stopping the iteration after each step, indicating that this limit slightly increases with each iteration. At $n=11$, the iteration has stopped. If we were to forcibly continue, we would have $\hat{a}_{12}=0.989528$ and $\hat{a}_{min,0,12}=0.990276$, which violates the iterative condition.

From Table \ref{tab:1}, it can be seen that a small portion of the reduction in extractable energy flows into the extracted energy, while the majority flows into the irreducible mass. Based on the energy utilization efficiency, it is indicated that $29.5\%$ of the change in extractable energy is converted into extracted energy, while $70.5\%$ is converted into irreducible mass. The results show that reducing the black hole's spin cannot extract all the corresponding rotational energy. This limitation arises from the nonlinear increase in irreducible mass. Furthermore, after the iteration terminates in the repetitive Penrose process, a remaining extractable energy of $0.195441M$ persists, indicating that a significant amount of energy remains to be extracted by other means. These results exhibit similarities with the case of Kerr black holes \cite{3}.

Next, we change the decay radius to $\hat{r}_d = 2.4$, keeping all other parameters the same as in Table \ref{tab:1}. The results are presented in Table \ref{tab:2}.
\begin{table}[htbp]
\centering
\small
\caption{Repetitive Penrose process with initial $\hat{A}=0.3$ and $\hat{r}_d=2.4$.}
\label{tab:2}
\begin{tabular}{ccccccccccc}
\hline
$n$ & $\frac{M_n}{M_0}$ & $\hat{a}_n$ & $\frac{\mu_{1,n}}{\mu_0}$ & $\hat{E}_{1,n}$ & $\frac{E_{extractable,n}}{M_0}$ & $\frac{E_{extracted,n}}{M_0}$ & $\frac{M_{{irr,n}}}{M_0}$ & $\xi_n$ & $\Xi_n$ &$\hat{a}_{min,0,n}$ \\
\hline
0 & 1.000000 & 1.000000 & 0.027659 & -0.077270 & 0.258751 & 0.000000 & 0.741249 & 0.000000 & 0.000000 & 0.166019 \\
1 & 0.999979 & 0.994667 & 0.027516 & -0.067348 & 0.212974 & 0.000021 & 0.787005 & 0.002137 & 0.000467 & 0.166052 \\
2 & 0.999960 & 0.989356 & 0.027374 & -0.057461 & 0.194108 & 0.000040 & 0.805852 & 0.001995 & 0.000617 & 0.166081 \\
3 & 0.999944 & 0.984067 & 0.027233 & -0.047609 & 0.179686 & 0.000056 & 0.820259 & 0.001854 & 0.000704 & 0.166105 \\
4 & 0.999931 & 0.978800 & 0.027093 & -0.037792 & 0.167569 & 0.000069 & 0.832363 & 0.001715 & 0.000752 & 0.166125 \\
5 & 0.999921 & 0.973554 & 0.026954 & -0.028009 & 0.156930 & 0.000079 & 0.842991 & 0.001577 & 0.000774 & 0.166141 \\
6 & 0.999914 & 0.968330 & 0.026817 & -0.018260 & 0.147344 & 0.000086 & 0.852569 & 0.001440 & 0.000775 & 0.166152 \\
7 & 0.999909 & 0.963128 & 0.026680 & -0.008545 & 0.138559 & 0.000091 & 0.861350 & 0.001304 & 0.000759 & 0.166160 \\
8 & 0.999906 & 0.957947 & 0.026544 & 0.001138 & 0.130409 & 0.000094 & 0.869497 & 0.001170 & 0.000729 & 0.166163 \\
\hline
\end{tabular}
\end{table}
From Table \ref{tab:2}, it can be observed that the iteration terminates because $\hat{E}_{1,8} > 0$ during the 8th iteration, making the next iteration impossible.

Next, we change the decay radius to $\hat{r}_d = 2.2$, keeping all other parameters the same as in Table \ref{tab:1}. We present the results in Table \ref{tab:3}.
\begin{table}[htbp]
\centering
\small
\caption{Repetitive Penrose process with initial $\hat{A}=0.3$ and $\hat{r}_d=2.2$.}
\label{tab:3}
\begin{tabular}{ccccccccccc}
\hline
$n$ & $\frac{M_n}{M_0}$ & $\hat{a}_n$ & $\frac{\mu_{1,n}}{\mu_0}$ & $\hat{E}_{1,n}$ & $\frac{E_{extractable,n}}{M_0}$ & $\frac{E_{extracted,n}}{M_0}$ & $\frac{M_{{irr,n}}}{M_0}$ & $\xi_n$ & $\Xi_n$ &$\hat{a}_{min,0,n}$ \\
\hline
0 & 1.000000 & 1.000000 & 0.026873 & -0.398875 & 0.258751 & 0.000000 & 0.741249 & 0.000000 & 0.000000 & 0.484488 \\
1 & 0.999893 & 0.994991 & 0.026729 & -0.390406 & 0.214372 & 0.000107 & 0.785521 & 0.010719 & 0.002415 & 0.484557 \\
2 & 0.999788 & 0.990002 & 0.026587 & -0.381968 & 0.196082 & 0.000212 & 0.803706 & 0.010577 & 0.003376 & 0.484624 \\
3 & 0.999687 & 0.985034 & 0.026445 & -0.373562 & 0.182102 & 0.000313 & 0.817585 & 0.010437 & 0.004085 & 0.484689 \\
4 & 0.999588 & 0.980087 & 0.026304 & -0.365186 & 0.170360 & 0.000412 & 0.829228 & 0.010297 & 0.004660 & 0.484753 \\
5 & 0.999492 & 0.975161 & 0.026165 & -0.356840 & 0.160053 & 0.000508 & 0.839439 & 0.010159 & 0.005146 & 0.484815 \\
6 & 0.999399 & 0.970254 & 0.026026 & -0.348525 & 0.150768 & 0.000601 & 0.848630 & 0.010022 & 0.005569 & 0.484875 \\
7 & 0.999308 & 0.965369 & 0.025889 & -0.340238 & 0.142262 & 0.000692 & 0.857046 & 0.009886 & 0.005941 & 0.484933 \\
8 & 0.999220 & 0.960503 & 0.025753 & -0.331981 & 0.134374 & 0.000780 & 0.864846 & 0.009751 & 0.006272 & 0.484990 \\
9 & 0.999134 & 0.955658 & 0.025617 & -0.323753 & 0.126992 & 0.000866 & 0.872142 & 0.009618 & 0.006570 & 0.485045 \\
10 & 0.999051 & 0.950833 & 0.025483 & -0.315553 & 0.120036 & 0.000949 & 0.879015 & 0.009485 & 0.006838 & 0.485098 \\
11 & 0.998971 & 0.946028 & 0.025349 & -0.307380 & 0.113445 & 0.001029 & 0.885526 & 0.009354 & 0.007081 & 0.485150 \\
12 & 0.998893 & 0.941244 & 0.025217 & -0.299236 & 0.107171 & 0.001107 & 0.891722 & 0.009224 & 0.007302 & 0.485200 \\
13 & 0.998818 & 0.936479 & 0.025086 & -0.291118 & 0.101177 & 0.001182 & 0.897641 & 0.009095 & 0.007503 & 0.485248 \\
14 & 0.998745 & 0.931735 & 0.024955 & -0.283028 & 0.095431 & 0.001255 & 0.903314 & 0.008967 & 0.007686 & 0.485295 \\
15 & 0.998674 & 0.927010 & 0.024825 & -0.274963 & 0.089908 & 0.001326 & 0.908766 & 0.008840 & 0.007853 & 0.485341 \\
16 & 0.998606 & 0.922305 & 0.024697 & -0.266925 & 0.084587 & 0.001394 & 0.914019 & 0.008714 & 0.008005 & 0.485384 \\
17 & 0.998540 & 0.917620 & 0.024569 & -0.258913 & 0.079450 & 0.001460 & 0.919090 & 0.008589 & 0.008144 & 0.485427 \\
18 & 0.998476 & 0.912954 & 0.024443 & -0.250927 & 0.074481 & 0.001524 & 0.923995 & 0.008465 & 0.008269 & 0.485468 \\
19 & 0.998415 & 0.908308 & 0.024317 & -0.242965 & 0.069668 & 0.001585 & 0.928747 & 0.008343 & 0.008383 & 0.485507 \\
20 & 0.998356 & 0.903682 & 0.024192 & -0.235028 & 0.064998 & 0.001644 & 0.933357 & 0.008221 & 0.008486 & 0.485545 \\
21 & 0.998299 & 0.899075 & 0.024068 & -0.227116 & 0.060463 & 0.001701 & 0.937836 & 0.008100 & 0.008579 & 0.485581 \\
22 & 0.998244 & 0.894488 & 0.023945 & -0.219228 & 0.056051 & 0.001756 & 0.942193 & 0.007980 & 0.008662 & 0.485616 \\
23 & 0.998192 & 0.889920 & 0.023823 & -0.211364 & 0.051757 & 0.001808 & 0.946435 & 0.007862 & 0.008736 & 0.485650 \\
24 & 0.998141 & 0.885371 & 0.023701 & -0.203523 & 0.047572 & 0.001859 & 0.950570 & 0.007744 & 0.008801 & 0.485682 \\
25 & 0.998093 & 0.880842 & 0.023581 & -0.195706 & 0.043490 & 0.001907 & 0.954604 & 0.007627 & 0.008858 & 0.485713 \\
26 & 0.998047 & 0.876331 & 0.023461 & -0.187911 & 0.039505 & 0.001953 & 0.958542 & 0.007511 & 0.008908 & 0.485743 \\
27 & 0.998003 & 0.871840 & 0.023343 & -0.180139 & 0.035613 & 0.001997 & 0.962390 & 0.007396 & 0.008950 & 0.485771 \\
28 & 0.997961 & 0.867368 & 0.023225 & -0.172390 & 0.031808 & 0.002039 & 0.966153 & 0.007282 & 0.008985 & 0.485798 \\
29 & 0.997921 & 0.862914 & 0.023108 & -0.164663 & 0.028085 & 0.002079 & 0.969835 & 0.007169 & 0.009014 & 0.485824 \\
30 & 0.997883 & 0.858479 & 0.022992 & -0.156957 & 0.024442 & 0.002117 & 0.973440 & 0.007057 & 0.009036 & 0.485848 \\
31 & 0.997847 & 0.854064 & 0.022876 & -0.149274 & 0.020875 & 0.002153 & 0.976972 & 0.006946 & 0.009052 & 0.485871 \\
32 & 0.997813 & 0.849666 & 0.022762 & -0.141611 & 0.017379 & 0.002187 & 0.980433 & 0.006836 & 0.009062 & 0.485893 \\
33 & 0.997780 & 0.845288 & 0.022648 & -0.133969 & 0.013953 & 0.002220 & 0.983828 & 0.006726 & 0.009067 & 0.485914 \\
34 & 0.997750 & 0.840928 & 0.022535 & -0.126349 & 0.010592 & 0.002250 & 0.987158 & 0.006618 & 0.009067 & 0.485933 \\
35 & 0.997722 & 0.836586 & 0.022423 & -0.118748 & 0.007295 & 0.002278 & 0.990427 & 0.006510 & 0.009061 & 0.485952 \\
36 & 0.997695 & 0.832263 & 0.022312 & -0.111168 & 0.004058 & 0.002305 & 0.993637 & 0.006403 & 0.009050 & 0.485969 \\
37 & 0.997670 & 0.827958 & 0.022201 & -0.103608 & 0.000881 & 0.002330 & 0.996790 & 0.006297 & 0.009035 & 0.485985 \\
\hline
\end{tabular}
\end{table}
According to Table \ref{tab:3}, the iteration stops at the 37th step. If we were to forcibly proceed to the 38th iteration, we would have $E_{extractable,38} = -0.002241 < 0$, which violates the iterative condition. From Table \ref{tab:3}, it can be seen that the irreducible mass reaches a remarkable $0.99679M$, leaving almost no remaining extractable energy. This indicates that under these parameters, nearly all of the extractable energy is converted into irreducible mass. The results shown in Table \ref{tab:3} differ from those of Kerr black holes, where, upon iteration termination, the remaining extractable energy is still relatively large, typically not less than $0.1M$ \cite{3}.

Finally, we change the decay radius to $\hat{r}_d = 1.06$, keeping all other parameters the same as in Table \ref{tab:1}. We present the results in Table \ref{tab:4}.
\begin{table}[htbp]
\centering
\small
\caption{Repetitive Penrose process with initial $\hat{A}=0.3$ and $\hat{r}_d=1.06$.}
\label{tab:4}
\begin{tabular}{ccccccccccc}
\hline
$n$ & $\frac{M_n}{M_0}$ & $\hat{a}_n$ & $\frac{\mu_{1,n}}{\mu_0}$ & $\hat{E}_{1,n}$ & $\frac{E_{extractable,n}}{M_0}$ & $\frac{E_{extracted,n}}{M_0}$ & $\frac{M_{{irr,n}}}{M_0}$ & $\xi_n$ & $\Xi_n$ &$\hat{a}_{min,0,n}$ \\
\hline
0 & 1.000000 & 1.000000 & 0.022403 & -9.704367 & 0.258751 & 0.000000 & 0.741249 & 0.000000 & 0.000000 & 0.999111 \\
1 & 0.997826 & 0.999990 & 0.022395 & -9.701363 & 0.256326 & 0.002174 & 0.741500 & 0.217407 & 0.896561 & 0.999111 \\
\hline
\end{tabular}
\end{table}
From Table \ref{tab:4}, it can be observed that the iteration stops after the first step. If we forcibly proceed to the second iteration, we find $\frac{M_{irr,2}}{M_0} = 0.740656 < \frac{M_{irr,1}}{M_0} = 0.741500$, meaning the irreducible mass decreases, which violates the iterative condition. Table \ref{tab:4} reveals a remarkable phenomenon: the energy utilization efficiency reaches $89.6\%$. This indicates that under these parameters, the reduction in extractable energy is primarily converted into extracted energy rather than irreducible mass. This result is significantly different from previous findings \cite{3,4,6}, where the energy utilization efficiency rarely exceeded $50\%$. Furthermore, if the decay radius is too small, such as $\hat{r}_d = 1-1.05$, even a single iteration cannot satisfy the conditions, making energy extraction impossible because the irreducible mass becomes imaginary in such cases.

\begin{figure}[!h]
  \centering
  \setlength{\tabcolsep}{2pt}
  \begin{tabular}{cc}
    \includegraphics[width=0.45\linewidth]{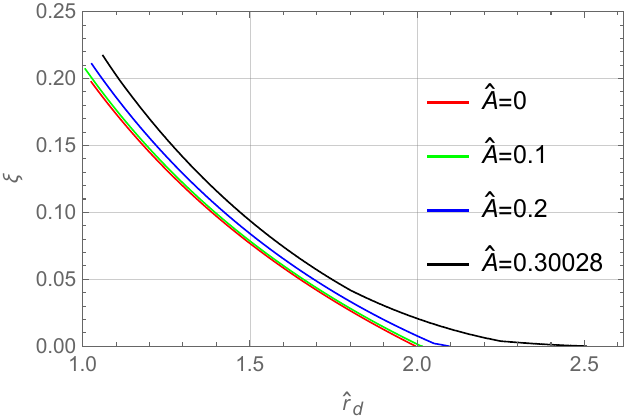} &
    \includegraphics[width=0.45\linewidth]{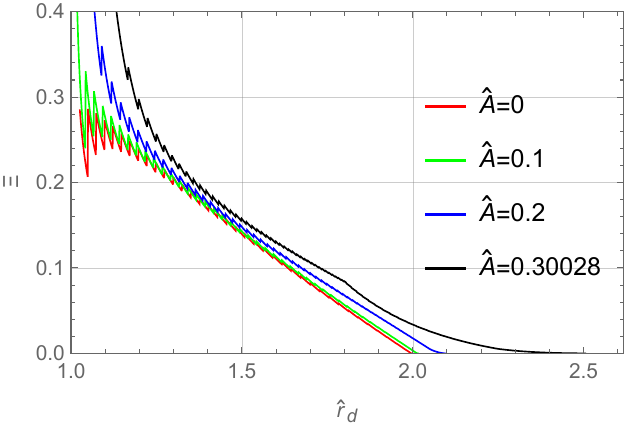} \\
    (a) & (b) \\[6pt]
    \includegraphics[width=0.45\linewidth]{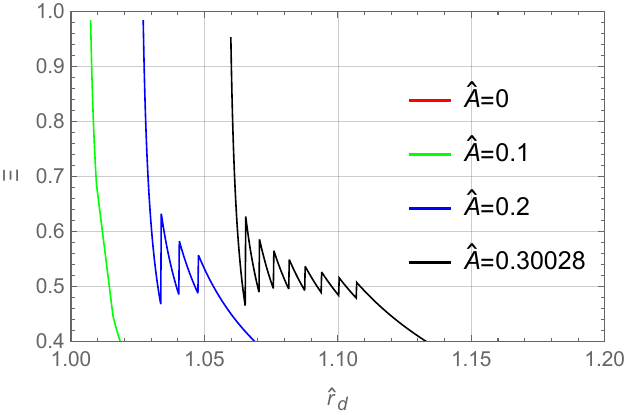} &
    \includegraphics[width=0.45\linewidth]{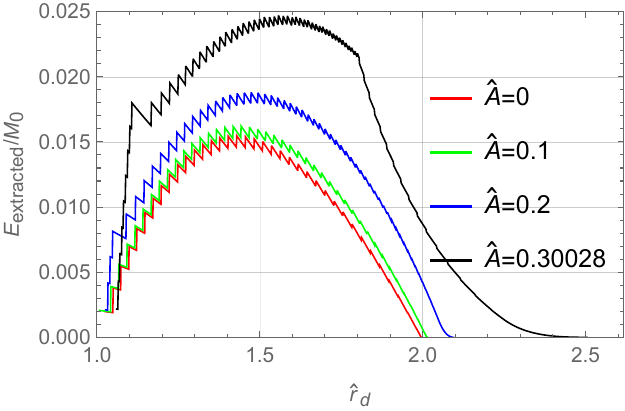} \\
    (c) & (d) \\[6pt]
    \includegraphics[width=0.45\linewidth]{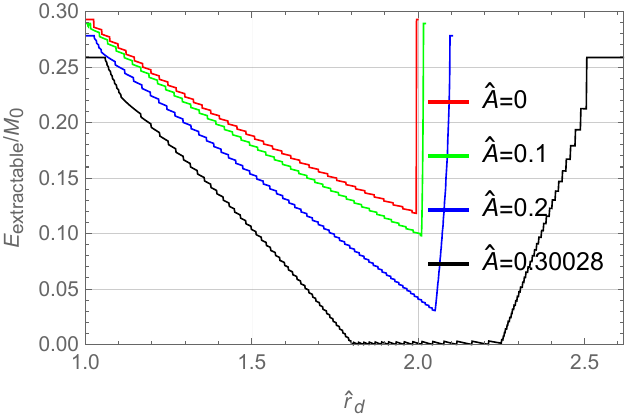} &
    \includegraphics[width=0.45\linewidth]{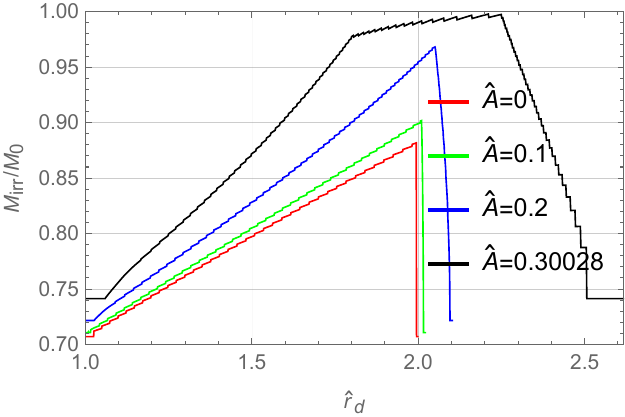} \\
    (e) & (f)
  \end{tabular}
  \caption{Under different initial $\hat{A}$ values, after the termination of the repetitive Penrose process: (a) the energy return on investment $\xi$; (b, c) the energy utilization efficiency $\Xi$; (d) the extracted energy $E_{extracted}/M_0$; (e) the extractable energy $E_{extractable}/M_0$; (f) the irreducible mass $\frac{M_{irr}}{M_0}$ as functions of the decay radius $\hat r_d$. Each oscillation in the curves corresponds to a different number of iterations, caused by the iterative conditions and reflecting the discrete nature of the process.}
  \label{fig:4}
\end{figure}

In Fig. \ref{fig:4}, we plot the variation with decay radius $\hat r_d$ of the energy return on investment $\xi$, the energy utilization efficiency $\Xi$, the extracted energy $E_{extracted}/M_0$, the extractable energy $E_{extractable}/M_0$, and the irreducible mass $\frac{M_{irr}}{M_0}$ after the repetitive Penrose process terminates, under different initial $\hat{A}$ values.

From panels (a), (b), and (d) of Fig. \ref{fig:4}, it can be observed that at the same decay radius, as the initial $\hat{A}$ increases, the values of the energy return on investment, the energy utilization efficiency, and the extracted energy almost all increase. An exception is a slight anomaly in the extracted energy for $\hat{A}=0.30028$ at lower decay radii. This indicates that, in the repetitive Penrose process, accelerating Kerr black holes possess stronger energy extraction capabilities compared to Kerr black holes. From panel (c) of Fig. \ref{fig:4}, it can be seen that at lower decay radii, the energy utilization efficiency of accelerating Kerr black holes can exceed $50\%$, suggesting that the reduced extractable energy is primarily converted into extracted energy rather than irreducible mass. This is a distinctive feature of accelerating Kerr black holes. From panel (e) of Fig. \ref{fig:4}, it is evident that when the initial value of the acceleration factor is large, the extractable energy can decrease to nearly zero. However, in such cases, the extractable energy is mainly converted into irreducible mass rather than extracted energy, as reflected in panel (f) where, for example, the irreducible mass increases to almost 1. Overall, the extracted energy still constitutes only a small fraction of the initial black hole mass, not exceeding $2.5\%$. This shows that the repetitive Penrose process, even when using the Ruffini process for energy extraction, still has significant limitations.

\section{Conclusion}
In this paper, under the optimal conditions for maximum energy extraction, we investigate energy extraction via the repetitive Penrose process in extremal accelerating Kerr black holes. First, we provide a brief review of accelerating Kerr black holes, including their horizons and ergospheres. Second, we introduce the fundamental equations of the Penrose process for accelerating Kerr black holes. Then, we describe the five iterative stopping conditions that the Penrose process must satisfy. In particular, we plot the variation of the lower spin limits for particles 0, 1, and 2 with the decay radius $\hat r_d$ under different $\hat{A}$ values, and compare the lower spin limits of the three particles, concluding that one of the lower spin limits for stopping the iteration is governed by particle 0. Finally, we present corresponding numerical results. Table \ref{tab:1} shows that the iteration stops due to the failure to meet the lower spin limit for particle 0; Table \ref{tab:2} shows that the iteration stops because $\hat{E}_{1} > 0$; Table \ref{tab:3} shows that the iteration stops due to $E_{extractable} < 0$; and Table \ref{tab:4} shows that the iteration stops because the irreducible mass begins to decrease.

Similar to previous conclusions, reducing the black hole's spin cannot extract all the corresponding rotational energy. This limitation arises from the nonlinear increase in irreducible mass. The differences lie in the fact that, at the same decay radius, as the initial $\hat{A}$ increases, the values of the energy return on investment, the energy utilization efficiency, and the extracted energy almost all increase. This demonstrates that, in the repetitive Penrose process, accelerating Kerr black holes possess stronger energy extraction capabilities compared to Kerr black holes. When the decay radius is relatively low, the energy utilization efficiency of accelerating Kerr black holes can exceed $50\%$, indicating that the reduction in extractable energy is primarily converted into extracted energy rather than irreducible mass. Furthermore, when the initial value of the acceleration factor is large, the extractable energy can decrease to nearly zero. These phenomena are entirely different from those observed in Kerr black holes.

\noindent {\bf Acknowledgments}

\noindent
This work is supported by the National Natural Science Foundation of China (Grants Nos. 12375043,
12575069 ).

\end{document}